\documentclass[12pt,a4paper]{article}

\catcode`\@=11
\@addtoreset{equation}{section}

\global\arraycolsep=2pt
\usepackage{mathrsfs,amsbsy,amssymb,latexsym,amsfonts,amsmath}
\usepackage{graphicx,color,cite}

\allowdisplaybreaks

\newcommand{\dd}{{\rm d}}
\newcommand{\gym}{g_{\text{YM}}}
\newcommand{\gs}{g_{\text{s}}}

\newcommand{\tf}{T_{\rm F} }
\newcommand{\diag}{{\rm{diag}} }

\newcommand{\rc}{r_{\rm c} }
\newcommand{\rh}{r_{\rm h} }

\newcommand{\const}{\text{const.} }

\usepackage[hypertex]{hyperref}

\begin{document}

\begin{flushright}
\parbox{4.2cm}
{KUNS-2449}
\end{flushright}

\vspace*{2cm}

\begin{center}
{\Large \bf Potential Analysis in Holographic Schwinger Effect}
\vspace*{2cm}\\
{\large Yoshiki Sato\footnote{E-mail:~yoshiki@gauge.scphys.kyoto-u.ac.jp} 
and 
Kentaroh Yoshida\footnote{E-mail:~kyoshida@gauge.scphys.kyoto-u.ac.jp} 
}
\end{center}

\vspace*{1cm}
\begin{center}
{\it Department of Physics, Kyoto University \\ 
Kyoto 606-8502, Japan} 
\end{center}

\vspace{1cm}

\begin{abstract}
We analyze electrostatic potentials in the holographic Schwinger effect. 
The potential barrier for the pair production is estimated by a static potential 
consisting of static mass energies, an electric potential from an  external electric-field, 
and the Coulomb potential between a particle and an antiparticle.  
Given that the Coulomb potential is supposed to be evaluated 
by the minimal surface attaching on the conformal boundary 
as usual, the critical field, where the potential barrier vanishes,  
exhibits a deviation of 30$\%$ from the one obtained from the Dirac-Born-Infeld (DBI) action.  
We reconsider this issue by reexamining the Coulomb-potential part, 
which is evaluated by the classical action of a string solution attaching on a probe D3-brane 
sitting at an intermediate position in the bulk AdS.   
Then the resulting critical-field completely agrees with the DBI result. This agreement gives rise to a strong  
support for the holographic scenario. We also discuss the finite-temperature case 
and the temperature-dependent critical-field also agrees with the DBI result. 
\end{abstract}

\thispagestyle{empty}
\setcounter{page}{0}

\newpage

\section{Introduction}

The Schwinger effect \cite{Schwinger} is known as a non-perturbative phenomenon 
in quantum electromagnetic dynamics (QED).  The virtual electron-position pairs can 
become real particles due to the presence of a strong electric-field. 
This phenomenon is not restricted to QED. Various kinds of strong external-fields can create 
a neutral pair of higher-dimensional objects such as strings and D-branes. 
Therefore, the Schwinger effect should be ubiquitous 
and would be a key ingredient to get a deeper understanding for the vacuum structure and 
non-perturbative aspects of string theory as well as quantum field theories. 

\medskip 

\begin{figure}[htbp]
\vspace*{0.5cm}
 \begin{center}
  \includegraphics[scale=0.8]{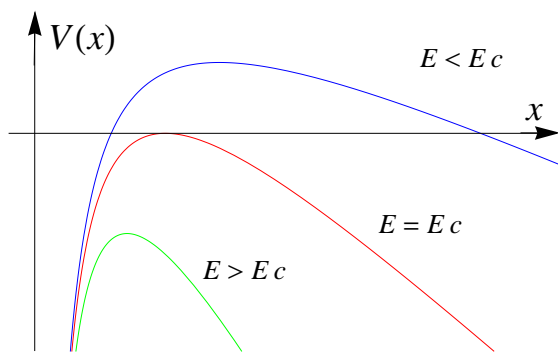}
  \end{center}
  \vspace*{-0.5cm}
  \caption{\footnotesize Shapes of the potential for some values of $E$\,. 
  The critical electric-field is denoted by $E_{\text{c}}$\,. 
\textcolor{blue}{The blue line} is the barrier with $E<E_{\text{c}}$\,. 
\textcolor{red}{The red line} with $E=E_{\text{c}}$\,. 
\textcolor{green}{The green line}  with $E>E_{\text{c}}$\,.}
\label{withint}
\end{figure}

To explain the issue we are concerned with, let us remember a potential analysis in considering 
the Schwinger effect in the context of QED. The potential barrier estimated by the static potential 
is related to a quantum tunneling process. 
For being real, the virtual electron-position pairs have to get a greater energy 
than the rest energy from an external electric field. 
Intuitively, when virtual pairs are separated by a distance $x$,
they gain an energy $eEx$ from the external electric-field. 
By including the Coulomb interaction between the particles, the total potential is estimated as 
\begin{equation}
V(x)=2m-eEx-\frac{\alpha _{\rm s}}{x}\,, 
\end{equation}
where $\alpha _{\rm s}$ is a fine-structure constant. The shape of the potential depends on the 
values of the external field as shown in Fig.\,\ref{withint}.  

\medskip 

It is easy to see that there are two kinds of instabilities of the vacuum. 
The first instability is the tunneling process for the potential barrier. 
This is the usual case. When the electric field is supposed to be small 
as denoted with the blue line in Fig.\,\ref{withint},  
the potential barrier is present and the pair production is described as a tunneling process. 
As a result, the production rate is exponentially suppressed.
The latter one occurs when the potential barrier vanishes.  
As the electric field becomes greater, the potential barrier decreases gradually.  
At last, it vanishes at a certain value of the electric field $E=E_{\text{c}}$ as depicted in Fig.\,\ref{withint}. 
Then no tunneling occurs and the production rate 
is not exponentially suppressed any more. 
The pair production is catastrophic and the vacuum becomes totally unstable. 
We refer to this electric field $E_{\text{c}}$ as the critical electric-field. 

\medskip 

If this potential analysis is true, the critical value of the external electric-field 
is supposed to be in QED. However, it might sound curious because such a critical value 
cannot be observed from the well-known results of the pair-production rate computed in QED. 
It is instructive to remember the formulas calculated by Schwinger \cite{Schwinger} and 
Affleck-Alvarez-Manton (AAM) \cite{AAM}, which are given by, respectively,
\footnote{The similar result is obtained for the rate of monopole pair production \cite{AM}.} 
\begin{align}
\text{Schwinger}:~~~&\Gamma \sim \exp \left( -\frac{\pi m^2}{eE}\right) 
	\qquad\qquad~ \text{for weak coupling and weak field}\,, 
\nonumber \\
\text{AAM}:~~~&\Gamma \sim \exp \left( -\frac{\pi m^2}{eE}+\frac{e^2}{4}\right) 
	\qquad \text{for arbitrary coupling and weak field}\,.  \nonumber 
\end{align}
In the Schwinger case there is no critical field trivially.   
In the AAM case, the exponential suppression vanishes when the electric field reaches $eE= (4\pi/e^2) m^2$\,. 
But the fine-structure constant $\alpha _{\rm s}=e^2/4\pi$ is around 1/137 in QED 
and the critical value does not satisfy the weak-field condition $eE \ll m^2$\,. 
Hence the existence of the critical value is not verified.  

\medskip 

From the above observations, the weak-field condition seems to be an obstacle 
to find out the critical field and hence it may be figured out 
if the production rate can be computed without this condition. 
Also, there exists the critical value of electric fields in string theory \cite{max1,max2} 
and this property may be regarded as a stringy nature. Hence the UV completion of the theory 
may be related to the critical behavior. 

\medskip 

The AdS/CFT correspondence \cite{M,GKP,W} provides a nice laboratory to test this argument. 
The holographic computation is done at strong coupling and there is no constraint 
for the values of external fields\footnote{As a matter of course, 
back reactions to the spacetime should be taken into account in the case of extremely-strong 
external-fields and there is always a limitation concerning the probe approximation.}.  
This point is a great advantage in comparison to the field-theoretical computation. 

\medskip 
 
Thus it is of great interesting to consider the Schwinger effect in a holographic setup and 
a series of studies in this direction was initiated by \cite{GSS,SZ}.  
An intriguing support for the existence of the critical electric-field
has recently been presented 
by Semenoff and Zarembo in a holographic computation \cite{SZ}.
The system is the $\mathcal{N}=4$ super Yang-Mills (SYM) theory and the $U(1)$ system 
is realized through the Higgs mechanism.  
The production rate of the fundamental particles (called the W-boson supermultiplet 
or  simply ``quarks'') with mass $m$\,,  
at large $N$ and large 't~Hooft coupling $\lambda=N\gym ^2$\,,  
has been computed as 
\begin{equation}
\Gamma \sim \exp \left[ -\frac{\sqrt{\lambda}}{2}
\left( \sqrt{\frac{E_{\text{c}} } {E}}-\sqrt{\frac{E } {E_{\text{c}}}} \, \right)^2\right]
 \,, \qquad E_{\text{c}}=\frac{2\pi m^2}{\sqrt{\lambda}}\,. 
\end{equation}
The critical field $E_{\text{c}}$ can really be seen here and it also agrees with the DBI result.  
A generalization to the case with magnetic fields has also been done \cite{BKR,SY}.

\medskip

On the other hand, a simple argument for the potential does not work in the above case, 
as noted in \cite{SZ}. The critical field can be estimated by using the Coulomb potential 
computed with the AdS/CFT prescription \cite{Wilson1,Wilson2}, 
\[
E_{\text{c}} \sim 0.70\,\frac{2\pi m^2}{\sqrt{\lambda}}\,. 
\]
This simple argument exhibits almost 30$\%$ deviation from both the string world-sheet result and the DBI result. 

\medskip 

In this paper we revisit this deviation and reconsider the critical field  
from the viewpoint of the potential analysis. We examine a static potential 
by evaluating the classical action of a string solution attaching on a probe D3-brane 
sitting at an intermediate position. Then the resulting critical-field completely agrees with the DBI result. 
We also discuss the finite-temperature case by following \cite{BKR} and show that 
the temperature-dependent critical-field also agrees with the DBI result. 

\medskip 

Note that our approach has an advantage in comparison to evaluating 
the expectation value of circular Wilson loop. 
For completing the analysis in \cite{SZ}, the existence of a negative eigenvalue 
has to be shown by evaluating one-loop determinants of fluctuations  
around the string world-sheet solution corresponding to the circular Wilson loop \cite{BCFM,DGO}. 
However, it has not been done yet because of subtleties associated with the regularization method  
(For attempts in this direction, see \cite{DGT,AmMa,KM}). 
Our approach has no connection with the difficulty and thus 
our result gives a strong support for the scenario proposed in \cite{SZ} 
from another perspective. 
 
\medskip 

The organization of this paper is as follows. 
In section 2 a potential analysis is done at zero temperature.  
The Coulomb-potential part is represented by a hypergeometric function. 
The total potential takes a bit complicated form but the critical value of the electric field  
can be evaluated numerically. The result completely agrees with the one obtained from the DBI action 
that describes the probe D3-brane.   
In section 3 the finite-temperature case is investigated with the planar AdS black hole background. 
The temperature-dependent critical-field is examined numerically again. 
This result also agrees with the one obtained from the DBI action. 
Section 4 is devoted to conclusion and discussion.

\section{Potential analysis at zero temperature}

\subsection{Setup} 

Let us introduce the setup that is necessary for our argument below. First of all, 
the metric of AdS$_5\times$S$^5$ in the Poincar\'e coordinates is given by 
\begin{equation}
\dd s^2=\frac{r^2}{L^2} \, \eta _{\mu \nu }\, \dd x^\mu \dd x^\nu
+\frac{L^2}{r^2}\, \dd r^2+L^2 \dd \Omega _5 ^2\,, 
\end{equation}
where $L$ is the AdS radius and $\dd \Omega_5^2$ describes S$^5$\,.  
The fundamental relation of AdS/CFT is $L^2/\alpha'=\sqrt{\lambda}$\,, where $\alpha'$ is 
the square of string scale $\ell_{\rm s}$ like $\alpha' \equiv \ell_{\rm s}^2$\,. 
The coordinates $x^{\mu}~(\mu=0,\ldots,3)$ describe a four-dimensional slice 
for each of the values of $r$\,. The metric $\eta_{\mu\nu}$ is given by 
\[
\eta_{\mu\nu} = \left\{
\begin{array}{ll}
\diag (-1,1,1,1) & \qquad \mbox{(in~Lorentzian)}  \\ 
\diag (1,1,1,1) & \qquad \mbox{(in~Euclidean)} 
\end{array} 
\right.\,. 
\]
Depending on the purpose, the signature of $\eta_{\mu\nu}$ is taken.  
The critical electric-field is argued with the DBI action in the Lorentzian signature, 
while the static potentials are computed in the conventional way with the Euclidean signature.  

\medskip 

Remember that a quark-antiquark potential is computed from the expectation value  
of a rectangular Wilson loop, where the loop is regarded as a trajectory of test particles 
with infinitely heavy mass. 
In the context of AdS/CFT, the expectation value at strong coupling can be computed 
with the area of a string world-sheet attaching to the Wilson loop \cite{Wilson1, Wilson2}. 
There are various ways to associate the test particles with the field-theory ingredients, 
depending the setup under consideration. Here the test particles are associated with 
the fundamental matter fields introduced by the Higgs mechanism in the $\mathcal{N}=4$ 
SYM via the gauge-symmetry breaking from $SU(N+1)$ to $SU(N) \times U(1)$\,. 

\medskip 

In the usual studies, the test particles are assumed to be infinitely heavy. 
However, this assumption is not appropriate for considering the Schwinger effect in a holographic way  
because the pair creation is severely suppressed due to the divergent mass.
Therefore the setup should be modified so that the production rate makes sense. 
A resolution has been proposed by Semenoff and Zarembo \cite{SZ}.  
It is to put a probe D3-brane at an intermediate position $r=r_0$ rather than close to the boundary.  
Then the mass $m$ becomes finite and depends on $r_0$ like 
\[
m=\tf\, r_0\,,
\]
where $\tf = 1/2\pi\alpha'$ is the fundamental string tension. 

\medskip 

\begin{figure}[htbp]
 \begin{center}
  \includegraphics[scale=0.5]{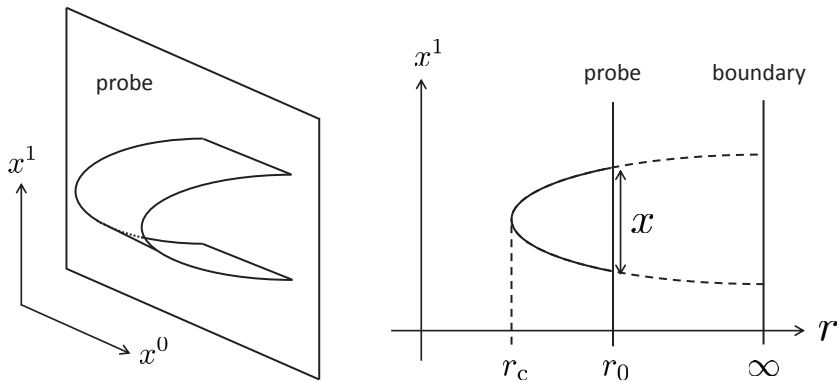}
%\vspace*{-0.5cm}
  \caption{\footnotesize The setup to compute the Coulomb potential. }
\label{figure}
 \end{center}
\end{figure}

Our purpose here is to compute the Coulomb potential from the area of the rectangular Wilson loop 
on the probe D3-brane. 
This computation can be done in a holographic way by evaluating the classical 
action of a string attaching on the probe D3-brane as depicted in Fig.\,\ref{figure}.  
This is a simple analog of the analysis on the circular Wilson-loop in \cite{SZ}.  

\subsection{Potential analysis}

The next step is to examine the classical solution of string world-sheet. 
We will work on the Euclidean signature here. Then the Nambu-Goto (NG) string action 
is given by 
\begin{eqnarray}
S &=& \tf \int \!\! \dd \tau \!\! \int\!\!  \dd \sigma \, \mathcal{L} \nonumber \\ 
&=& \tf \int \!\! \dd \tau \!\! \int\!\!  \dd \sigma \, \sqrt{\det G_{ab}}\,, \qquad  G_{ab} \equiv  
\frac{\partial x^{\mu}}{\partial \sigma ^a}\frac{\partial x^{\nu}}{\partial \sigma ^b}\, g_{\mu \nu}\,. 
\end{eqnarray}
Here $G_{ab}~(a,b=0,1)$ is the induced metric and 
the world-sheet coordinates are $\sigma^a =(\tau,\sigma)$\,. 
For our purpose, it is convenient to take the static gauge, 
\begin{equation}
x^0=\tau\, ,\qquad x^1=\sigma\,.
\end{equation}
For the classical solution, suppose that the radial direction depends only on $\sigma$\,, 
\begin{eqnarray}
r = r(\sigma)\,,
\end{eqnarray} 
and that the string solution is sitting at a point on S$^5$\,. 
The second condition means that a parameter $\theta ^I$ in a Wilson loop in $\mathcal{N}=4$ SYM 
is a constant $\theta _0^I$\,. 

\medskip 

Under this ansatz,  $\mathcal{L}$
is expressed as 
\begin{equation}
{\cal L} = \sqrt{\left(\frac{\dd r}{\dd \sigma}\right)^2+\frac{r^4}{L^4}}
\end{equation}
and does not depend on $\sigma$ explicitly. Hence the following quantity 
\begin{eqnarray}
\frac{\partial {\cal L}}{\partial (\partial _\sigma r)}\, \partial _\sigma r-{\cal L} 
\label{conserved}
\end{eqnarray}
is conserved and the relation 
\begin{align}
-\frac{r^4/L^4}{\displaystyle \sqrt{\left(\frac{\dd r}{\dd \sigma}\right)^2+\frac{r^4}{L^4}} }=\const 
\end{align}
is satisfied. 
By imposing the boundary condition at $\sigma=0$\,,  
\begin{equation}
\frac{\dd r}{\dd \sigma} =0\,, \qquad  r=\rc~~~~(\rc < r_0)\,, 
\label{bc}
\end{equation}
a differential equation is derived,  
\begin{align}
\frac{\dd r}{\dd \sigma}=\frac{r^2}{L^2}\sqrt{\frac{r^4}{\rc ^4}-1}\,. \label{4.28}
\end{align}
By introducing a new coordinate $y=r/\rc$ 
and integrating \eqref{4.28}\,, the separation length $x$ of test particles on the probe brane is represented by  
\begin{align}
x&=\frac{2L^2}{\rc}\int_1^{r_0/\rc}\frac{\dd y}{y^2\sqrt{y^4-1}} \notag \\ 
&=\frac{2L^2}{r_0} \left[ \frac{\sqrt{\pi }\, \Gamma (3/4) }{a\,\Gamma (1/4)}
-\frac{a^2}{3}\, {}_2F_1 \left(\frac{1}{2},\frac{3}{4},\frac{7}{4},a^4\right)\right]\,.
\end{align}
Here a dimensionless parameter $a$ is defined as 
\[
a \equiv \frac{\rc}{r_0}\,.
\]
This parameter measures the position of the tip of the classical solution 
with respect to the location of the probe D3-brane. 

\medskip 

\begin{figure}[tbp]
\begin{center}
\includegraphics[scale=0.8]{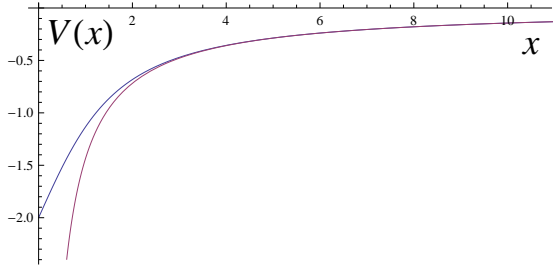}
\end{center}
\vspace*{-0.5cm}
\caption{\footnotesize The modified Coulomb potential versus the usual one. 
The modified potential is finite even at $x=0$\,, 
while at long distances these exhibit the same behavior. 
Note that the static energy is subtracted. 
\label{potential:fig}}
\end{figure}

By following the setup in Fig.\,\ref{figure}, 
the sum of the Coulomb potential (CP) and static energy (SE) 
can be estimated in a modified form like\footnote{
The potential with a specific electric field is computed in \cite{KL}.}
\begin{align}
V_{\rm CP+SE}&=2\tf\int^{x/2}_0\!\!\dd r\,\mathcal{L} =
2\tf\, \rc \int_{1}^{r_0/\rc} \!\dd y \, \frac{y^2}{\sqrt{y^4-1}} \notag \\
	&=2\tf\, r_0 \, \left[- \frac{a\,\sqrt{\pi }\, \Gamma (3/4) }{\Gamma (1/4)} 
	+ {}_2F_1 \left(-\frac{1}{4},\frac{1}{2},\frac{3}{4},a^4\right)\right]\,. \label{C-p}
\end{align}
The Coulomb-potential part is plotted in Fig.\,\ref{potential:fig}.  

\medskip 

Let us see the behavior of the potential (\ref{C-p}) in the $a\to 0$ limit,  
which is realized by putting the probe D3-brane off to the boundary 
(i.e., $r_0 \to \infty$ with $r_{\text{c}}$ fixed). With the help of the relation, 
\begin{eqnarray}
\lim_{a\to 0}{}_2F_1\left(-\frac{1}{4},\frac{1}{2},\frac{3}{4},a^4\right) = 1\,, 
\end{eqnarray} 
the second term in (\ref{C-p}) is reduced to the static energy of the test particles 
with infinitely heavy mass. 
Then $x$ is proportional to $1/a$ and thus the first term in (\ref{C-p}) gives rise to the well-known Coulomb potential.

\medskip 

When $a=1$\,, by using Gauss's Hypergeometric Theorem, the second term in (\ref{C-p}) can be rewritten as 
\begin{eqnarray}
{}_2F_1\left(-\frac{1}{4},\frac{1}{2},\frac{3}{4},1\right) 
= \frac{\Gamma(3/4)\,\Gamma(1/2)}{\Gamma(1)\,\Gamma(1/4)} 
= \frac{\sqrt{\pi}\,\Gamma(3/4)}{\Gamma(1/4)}\,.
\end{eqnarray}
Therefore the potential (\ref{C-p}) vanishes at $a=1$\,.  
This is also obvious from the integral expression in (\ref{C-p}). 

\medskip 

What is the physical meaning of the second term in (\ref{C-p})? 
The potential (\ref{C-p}) is finite at $x=0$ 
while the usual Coulomb potential diverges there (See Fig.\,\ref{potential:fig}).  
The finiteness at $x=0$ reminds us the Coulomb potential 
in non-linear electrodynamics\footnote{For this point,  as an example, see Section 7 of \cite{NL}.}. 
Indeed, the potential (\ref{C-p}) is interpreted as the potential (+ static energy) 
in the DBI theory on the probe D3-brane. 
The short-range behavior is described by a linear potential\footnote{
For a D7-brane case, see \cite{KMMW}.} 
\begin{equation}
V_{\rm CP+SE}\simeq \tf \left(\frac{r_0}{L}\right)^2x\,.
\end{equation}
It is rather obvious that the short-distance behavior of the potential has been modified 
because  we have cut out the UV part of the classical solution (near the boundary).

\medskip 

\begin{figure}[tbp]
\begin{center}
\includegraphics[scale=0.35]{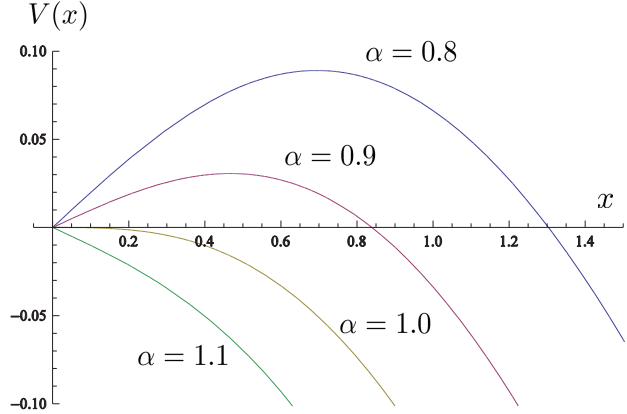}
\end{center}
\vspace*{-1.5cm}
\caption{\footnotesize The shapes of the potential at zero temperature 
for $\alpha=0.8,\,0.9,\,1.0$ and $1.1$ are plotted from top to bottom, 
where $\tf r_0=L^2/r_0=1$\,. The potential barrier vanishes for $\alpha \geq 1.0$ and the critical field is given by $\alpha=1.0$\,. 
Note that the behavior around $x=0$ is different from the one depicted in Fig.\ \ref{withint}. 
\label{zero:fig}}
\end{figure}

Let us next consider the total static-potential by turning on an electric field $E$ along the $x^1$-direction. 
It is convenient to introduce a dimensionless electric-field $\alpha$\,, 
\begin{eqnarray}
\alpha \equiv \frac{E}{E_{\rm c}}\,, \qquad  E_{\rm c} = \tf\, \frac{r_0^2}{L^2} 
\end{eqnarray}
by measuring $E$ in a unit of the critical field $E_{\rm c}$ obtained from the DBI argument. 
Together with the electrostatic potential associated with $E$\,, 
the total potential $V_{\rm tot}$ is given by  
\begin{align}
V_{\rm tot} &=V_{\rm CP+SE}-Ex  \notag \\
\begin{split}
	&=2 \tf\, r_0 \left\{ - \frac{a\sqrt{\pi }\, \Gamma (3/4) }{\Gamma (1/4)} +{}_2F_1
		 \left(-\frac{1}{4},\frac{1}{2},\frac{3}{4},a^4\right) \right. 
 \\ & \qquad	\qquad  \quad \left. 	
-\frac{\alpha}{a}\left[ \frac{\sqrt{\pi }\, \Gamma (3/4) }{\Gamma (1/4)}
-\frac{a^3}{3}\, {}_2F_1 \left(\frac{1}{2},\frac{3}{4},\frac{7}{4},a^4\right)\right]\right\}\,. 
\end{split}
\label{tot}
\end{align}
It seems difficult to figure out the critical value of the electric field analytically from the expression (\ref{tot}). 
But it is an easy task to estimate its shape numerically as a function of $x$\,. 
The shapes for $\alpha=0.8,\,0.9,\,1.0$ and $1.1$ are plotted in Fig.\,\ref{zero:fig}. 
The potential barrier vanishes for $\alpha \geq 1.0$ and the critical field is $\alpha=1.0$\,. 
Thus the potential (\ref{tot}) leads to the critical field that is identical with the DBI result\footnote{
For an analytic proof that the potential barrier vanishes at $E=E_{\rm c}$\,, see \cite{SY2}.}. 

\section{Potential analysis at finite temperature}

\subsection{Setup}

The next subject is to consider a generalization to the finite-temperature case by following \cite{BKR}. 
We work on the Lorentzian signature here.  
The metric of an AdS planar black hole is given by \cite{HS}   
\begin{equation}
\dd s^2=-\frac{r^2}{L^2}\left(1-\frac{\rh ^4}{r^4}\right) \dd t^2
+\frac{L^2}{r^2}\left(1-\frac{\rh ^4}{r^4}\right)^{-1}\dd r^2+\frac{r^2}{L^2} \,
\sum_{i=1}^3(\dd x^i)^2+L^2\dd \Omega _5^2\,.
\label{bh}
\end{equation}
The coordinates $x^i~(i=1,2,3)$ denote the spatial directions on a four-dimensional slice 
described with a fixed value of the radial coordinate $r$\,. 
The horizon is located at $r=\rh$ and 
the temperature of the black hole   
\begin{equation}
T=\frac{\rh}{\pi L^2} \label{ondo}
\end{equation}
is identified with that of the gauge-theory dual. 

\medskip 

Let us consider the DBI action of a probe D3-brane on the black-hole background (\ref{bh}), 
including a constant, world-volume electric-field $E$\,. The probe D3-brane is located at $r=r_0$ 
and the following relation is assumed below:
\[
\rh < \rc < r_0\,.  
\]
The classical action is written into the form,   
\begin{align}
S &=-T_{\text{D3}} \int\!\! \dd ^4 x \, \sqrt{-\det (g_{\mu \nu}+{\cal F}_{\mu \nu})}  \notag \\
&=-T_{\text{D3}}\frac{r_0^4}{L^4}\sqrt{1-\frac{\rh^4}{r_0^4}}\int\!\! \dd ^4 x \, 
\sqrt{1-\frac{(2\pi \alpha ')^2L^4}{ r_0^4(1-\rh^4/r_0^4)}E^2}\,, 
\end{align}
where $T_{\text{D3}}$ is the D3-brane tension given by  
\[
T_{\text{D3}} = \frac{1}{\gs (2\pi)^3\alpha '^2}\,.
\]
The DBI action becomes ill-defined for $E > E_{\text{c}}$\,, where $E_{\text{c}}$ is the critical value given by   
\begin{equation}
E_{\text{c}} =\tf \frac{r_0^2}{L^2}\,\sqrt{1-\frac{\rh^4}{r_0^4}}\,.  
\label{f-c} 
\end{equation}
One can see that $E_{\text{c}}$ implicitly depends on the temperature. 

\medskip 

It is instructive to rewrite the mass of the fundamental matter in terms of the gauge-theory  parameters  
and see the temperature dependence of $E_{\text{c}}$\,. 
The induced metric on the string world-sheet is
\begin{eqnarray}
g_{ab}=\diag \left(
-\frac{r^2}{L^2}\left(1-\rh^4/r^4\right) , \frac{L^2}{r^2}(1-\rh^4/r^4)^{-1} 
\right) \label{induced-ft}
\end{eqnarray}
and then the mass is represented by 
\begin{equation}
m=\tf \int_{\rh}^{r_0}\!\dd r \, \sqrt{-\det g_{ab}}\, = \tf \left(r_0-\rh\right)\,. 
\end{equation}
Thus the mass also depends on the temperature. It is notable that the temperature dependence comes from 
the integration range rather than the induced metric (\ref{induced-ft}). 

\medskip 

The temperature-dependent mass $m(T)$ is written into the well-known form, 
\begin{eqnarray}
m(T) = m(T=0) + \Delta m(T) \equiv m(T=0) - \frac{\sqrt{\lambda}}{2}T\,.
\end{eqnarray}
Then $E_{\text{c}}$ is also expressed in terms of the gauge-theory parameters as  
\begin{eqnarray}
E_{\text{c}}(T) &=& \frac{2\pi m^2}{\sqrt{\lambda}}\sqrt{
\left(1- \frac{\Delta m(T)}{m}\right)^4 - \left(\frac{\Delta m(T)}{m}\right)^4
} \nonumber \\ 
&\simeq& \frac{2\pi m^2}{\sqrt{\lambda}} \left(1-2 \frac{\Delta m(T)}{m}\right) 
+\mathcal{O}\left(\frac{\Delta m(T)}{m}\right)
\qquad (m \gg \sqrt{\lambda}\,T)
\,. 
\label{f-c-2}
\end{eqnarray}
It is worth noting that $E_{\text{c}}$ is well defined for arbitrary values of $T$\,. 

\subsection{Potential analysis}

The analysis on the classical solution is similar to the zero-temperature case. 
We work on the Euclidean signature here. 
The ansatz for the solution 
is also not changed. The induced metric $G_{ab}$ is slightly modified and
the NG action is rewritten as  
\begin{align}
S=
\tf \int\!\! \dd \tau\! \int
\!\! \dd \sigma \, {\cal L}\,, \qquad 
{\cal L} = \sqrt{\left(\frac{\dd r}{\dd \sigma}\right)^2 +\frac{r^4}{L^4}\left( 1-\frac{\rh^4}{r^4}\right)}\,.  
\end{align}
The quantity in (\ref{conserved}) 
is conserved again and the relation 
\begin{align}
-\frac{r^4/L^4}{\displaystyle \sqrt{\left(\frac{\dd r}{\dd \sigma}\right)^2+\frac{r^4}{L^4}
\left( 1-\frac{\rh^4}{r^4}\right)} }\left( 1-\frac{\rh^4}{r^4}\right)=\const
\end{align}
is satisfied. 
By imposing the boundary condition (\ref{bc}) at $\sigma=0$ again, we obtain the following differential 
equation, 
\begin{align}
\frac{\dd r}{\dd \sigma}=\frac{r^2}{L^2}
\sqrt{\left( 1-\frac{\rh^4}{r^4}\right)\frac{r^4-\rc^4}{\rc^4-\rh^4}}\,. \label{df-f}
\end{align}
Integrating (\ref{df-f}), the distance $x$ between the test particles is represented 
by the following integral, 
\begin{equation}
x
=\frac{2L^2}{r_0}\frac{1}{a}\sqrt{1-\frac{b^4}{a^4}}\int_1^{1/a} \! \frac{\dd y}{\sqrt{(y^4-1)(y^4-b^4/a^4)}} \,.
\end{equation}
Here we have introduced the following dimensionless quantities: 
\[
y=\frac{r}{\rc} \,, \qquad a=\frac{\rc}{r_0} \,, \qquad b=\frac{\rh}{r_0}\,. 
\]
The sum of the Coulomb potential between the fundamental fields and the static energy 
is given by 
\begin{equation}
V_{\rm CP+SE} =2\tf \int_{0}^{x/2}\! \dd \sigma \, {\cal L}
	=2\tf\, r_0\,a \int_{1}^{1/a}\! \dd y\, \sqrt{ \frac{y^4-b^4/a^4}{y^4-1}}\,. 
	\label{C-p-f}
\end{equation}

\medskip 

\begin{figure}[htbp]
\begin{center}
\includegraphics[scale=0.35]{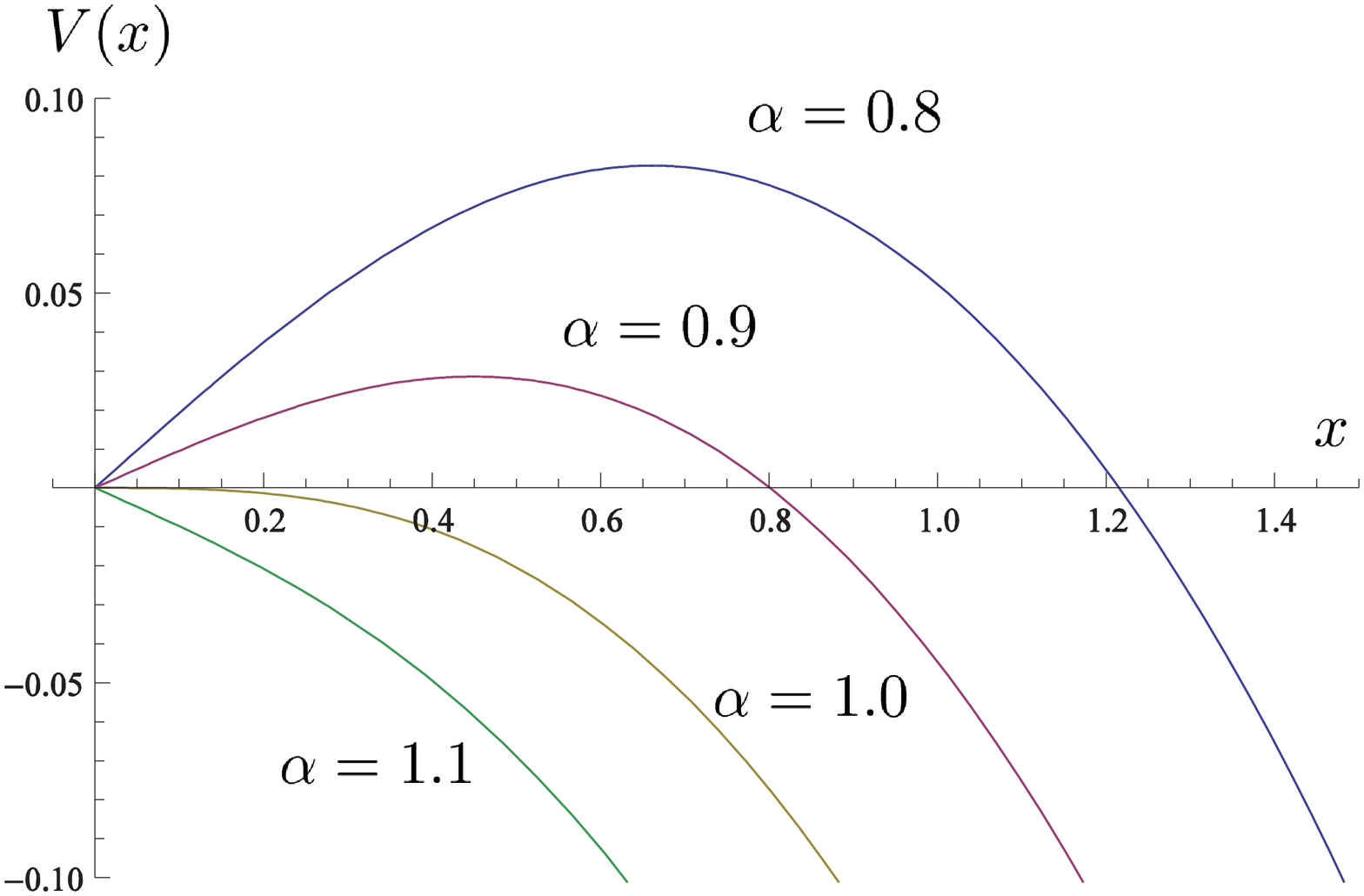}
\end{center}
\vspace*{-1.5cm}
\caption{\footnotesize The shapes of the potential with a fixed temperature ($b=0.5$) are plotted 
for $\alpha=0.8,\,0.9,\,1.0$ and $1.1$ from top to bottom, where $\tf r_0=L^2/r_0=1$. 
The potential barrier vanishes for $\alpha\geq 1.0$ and the critical field is $\alpha=1.0$\,. \label{finite:fig}}
\end{figure}

Let us next turn on an electric field $E$ along the $x^1$-direction. 
It can be measured with respect to the critical value (\ref{f-c}) 
by using a dimensionless parameter $\alpha$ like 
\[
\alpha  = \frac{E}{E_{\rm c}(T)}\,,  
\]
where $E_{\rm c}(T)$ is the critical field determined by the DBI argument and is given in (\ref{f-c}).
Then the total potential $V_{\rm tot}$ is given by  
\begin{align}
V_{\rm tot} &= V_{\rm CP+SE} - E x  \notag \\
\begin{split}
&=2\tf\, r_0\left[ ~a \int_{1}^{1/a}\! \dd y\, \sqrt{ \frac{y^4-b^4/a^4}{y^4-1}} \right.  \\ 
& \qquad \qquad 
\left. -\frac{1}{a}\sqrt{(1-b^4)\left(1-\frac{b^4}{a^4}\right)}\int_1^{1/a}\! 
\frac{\dd y}{\sqrt{(y^4-1)(y^4-b^4/a^4)}}\right]\,. 
\end{split}
\label{f-pot}
\end{align}
What we can do at most is to evaluate the potential shape (\ref{f-pot}) numerically again.  
The shapes of the potential (\ref{f-pot}) with a fixed temperature $b=0.5$ 
are plotted for the values $\alpha=0.8,\,0.9,\,1.0$ and $1.1$ in Fig.\,\ref{finite:fig}. 
The potential barrier vanishes for $\alpha \geq 1.0$ and the critical field is $\alpha=1.0$ again.  
Thus the critical field agrees with the DBI result.  

\medskip 

\begin{figure}[htbp]
\vspace*{0.5cm}
\begin{center}
\begin{tabular}{cc}
  \includegraphics[scale=.7]{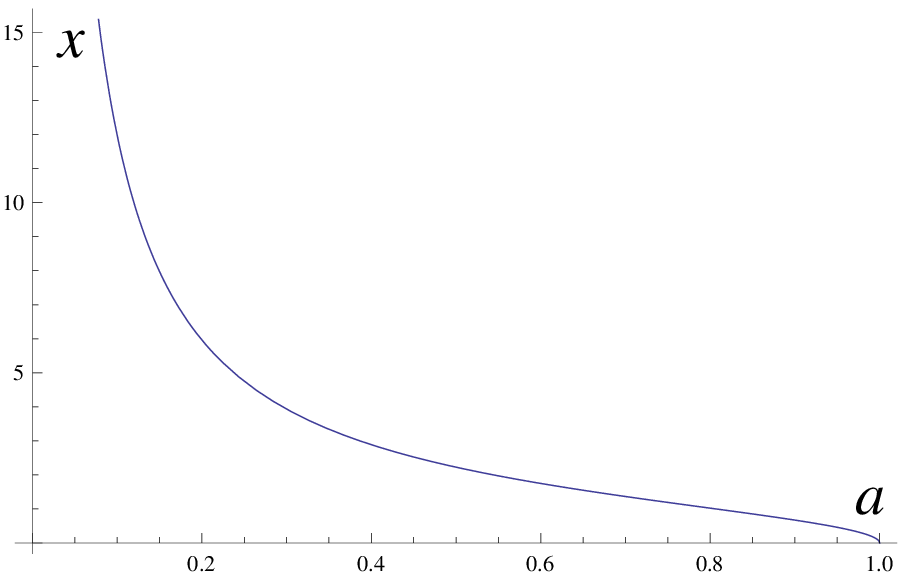} &  \qquad \includegraphics[scale=.7]{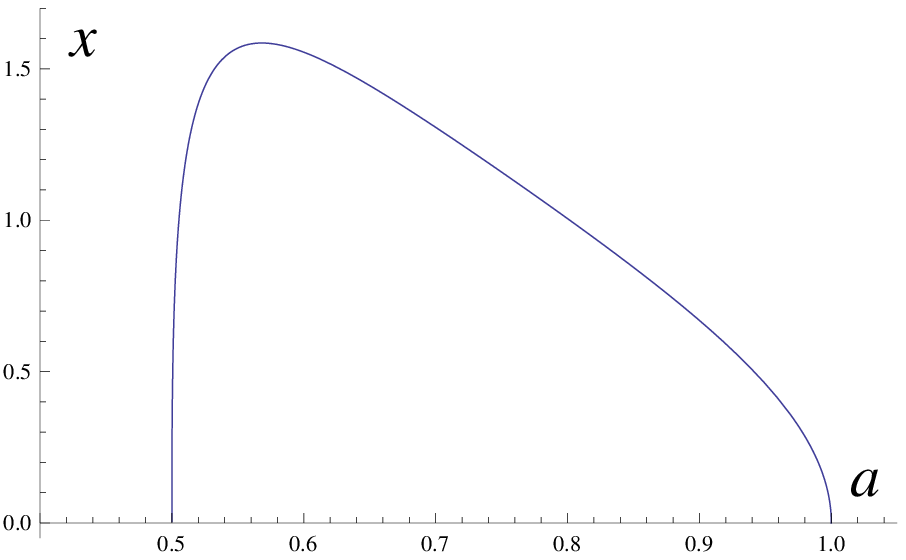} \\ 
  {\footnotesize $b=0$} & {\footnotesize $b=0.5$} 
\end{tabular}
\end{center}
\vspace*{-0.5cm}
\caption{\footnotesize The distance $x$ is plotted against $a$ for $b=0$ and $b=0.5$\,. 
\label{a-x:fig}}
\end{figure}

It should be remarked here that there is a degeneracy between the distance $x$ and the auxiliary parameter $a$ 
at finite temperature (while no degeneracy at zero temperature). 
The distance $x$ is plotted against $a$ for $b=0$ and $0.5$ in Fig.\,\ref{a-x:fig}. The value of $x$ is uniquely determined 
for a given value of $a$ at zero temperature ($b=0$), 
while a single value of $x$ corresponds to two values of $a$ at finite temperature. 
Therefore the region of $a$ has to be restricted for numerical computations. In fact, the branch with smaller values of $a$ means 
the region close to the horizon and describes another configuration of 
string world-sheet, a pair of straight Wilson lines, as discussed in \cite{RTY}. 
Thus the branch with larger values of $a$ has been taken for the present analysis. 
For Fig.\,\ref{finite:fig}\,, the range between 0.6 and 1.0 is utilized. 

\section{Conclusion and discussion}

We have analyzed electrostatic potentials in the holographic Schwinger effect 
by evaluating the classical action of a string solution attaching 
on a probe D3-brane sitting at an intermediate position in the bulk AdS. 
It has been shown numerically that 
the resulting critical-field agrees with the DBI result and the 30$\%$ deviation 
denoted in Introduction has been resolved. 
This agreement gives a strong support for the scenario proposed in \cite{SZ}.  
The finite-temperature case has also been considered with the planar AdS black hole. 
We have shown numerically that the temperature-dependent critical-field also agrees with the DBI result. 

\medskip 

It is also nice to consider the holographic Schwinger effect in the D3/D7-branes system. 
The pair creation of quark and antiquark and its time evolution are 
discussed in \cite{D7-1,D7-2} (For a review on quark dynamics in the holographic approach, 
for example, see \cite{review}). It would not be difficult to follow our approach in the D3/D7-branes 
system and it is interesting to consider the relation between the critical field and 
the time evolution of quark-antiquark pairs.   
We hope that we could report on this issue in the near future. 

\section*{Acknowledgments}

We would like to thank Io Kawaguchi and Koji Hashimoto for useful discussions. 
This work was also supported in part by the Grant-in-Aid 
for the Global COE Program ``The Next Generation of Physics, Spun 
from Universality and Emergence'' from MEXT, Japan.

\end{document}